\documentclass[epj]{svjour}
\usepackage{graphicx}

\usepackage{amssymb}
\usepackage{amsmath}

\usepackage{cuted}

\usepackage[colorlinks]{hyperref}

\AtBeginDocument{%
    \hypersetup{
      urlcolor     = blue, 
      linkcolor    = black, 
      citecolor   = blue, 
      pdfborder={0 0 0},
      urlbordercolor={1 1 1},
      citebordercolor={1 1 1}
    }
    }

\begin{document}

\hyphenation{coun-ter cor-res-pon-din-gly e-xam-ple
co-o-pe-ra-tion ex-pe-ri-men-tal mo-no-chro-ma-tic re-so-lu-tion
con-si-de-ring lo-ca-li-za-tion ma-xi-mum mo-di-fies}

\title{The quantum optical description of a Fabry-Perot interferometer and the prediction of an antibunching effect}

\author{Stefan~Ataman\inst{1}
\thanks{\emph{Present address:} ECE Paris, 37 Quai de Grenelle, Batiment Pollux, 75015 Paris, France}%
}                     
%
%
\institute{ECE Paris, \email{ataman@ece.fr}}
\date{Received: date / Revised version: date}

%
\abstract{In this paper we describe a Fabry-Perot interferometer
in the language of quantum optics. We go on to model the
Fabry-Perot interferometer as a beam splitter having frequency
dependent transmissivity and reflectivity coefficients. The
antibunching, a totally non-classical effect, is to be expected if
one excites this interferometer with carefully frequency-selected
single photons arriving simultaneously from opposite directions.
Contrary to a normal beam splitter, even slightly different
frequency single photons should be able to show this effect, as
long as the photon counters are not frequency selective.
%
\PACS{
      {42.50.-p}{Quantum optics}
     } 
} 

\titlerunning{The QO description of a Fabry-Perot interferometer and the prediction of an antibunching effect}
\authorrunning{S. Ataman}

\maketitle

%

\section{Introduction}

Ever since its introduction, more than a century ago, the
Fabry-Perot (FP) interferometer \cite{FP1899} has been ubiquitous
in spectroscopy and high resolution interferometry. A couple of
decades later, laser resonators \cite{Hod97} adopted its working
principle. Other FP applications include gravitational wave
detectors \cite{Tho80} and, more recently, for the needs of cavity
quantum electrodynamics (CQED), ultra-high finesse superconducting
\cite{Kuh07} and fiber high finesse \cite{Hun10} Fabry-Perot
cavities.

The FP interferometer is an optical device composed of two highly
reflecting mirrors and its classical description \cite{Hod97}
shows an optical device having extremely sharp variations going
back an forth between total transmission and total reflection as
the spacing between the two mirrors (or the incident light
frequency) is varied. There is a discrete set of frequencies,
where, whatever the transmissivity of the individual mirrors is,
the FP interferometer shows a total transmission. This effect can
be explained classically, as well as quantum mechanically.

The dynamic behavior of a FP interferometer has been investigated
by Lawrence et al. in \cite{Law99}, with application in the
context of Laser Gravitational-Wave Observatory (LIGO). It has
been shown that a different behavior is observed in the case of a
vibrational mirror and the application of a frequency modulated
signal to the FP interferometer. The extension of the model and
experimental results of Lawrence et al. was done by Rohde et al.
\cite{Roh02}, where the modification of the natural decay time of
atoms trapped inside the cavity was demonstrated.

Interesting experiments \cite{Fra14,Het11} showed that a single
atom can behave like an optical FP cavity under certain
conditions, thus bringing a traditionally macroscopic
interferometer to the atomic scale. In \cite{Sri14}, Srivathsan et
al. showed that a FP interferometer is able to reverse the
exponentially-falling into an exponentially-rising temporal
envelope for a heralded single photon, a feature with potential
interest in quantum information processing \cite{Bou00}. A recent
proposal by Sun \cite{Sun14} uses FP interferometers to precisely
resolve the continuous variable time-energy entanglement.

Non-classical states of light have been used for decades and the
most popular source of such states is the so-called spontaneous
parametric down-conversion (SPDC) \cite{Bur70,Kly67}, where an
incident pump photon, through a non-linear optical process,
generates two output photons, generally called ``signal'' and
``idler''.

One non-classical effect is the so-called ``HOM'' or anti-bunching
effect \cite{HOM87}, experimentally demonstrated by Hong, Ou and
Mandel. It manifests itself in a dip in the coincidence rate at
the output of a beam splitter, when two single-photon Fock states
impinge simultaneously at its inputs.

The quantum optical description of a FP interferometer is
typically done by quantizing the classical field modes
\cite{Ley87}. In this paper we start directly from a quantum
optical model of the FP interferometer. Being an optical lossless
component, it is expected that it will show a $SU(2)$ symmetry
\cite{Yur86,Cam89}.

Therefore, using a graphical method introduced in \cite{Ata14b}
and already used in \cite{Ata15a}, we describe the input-output
field operator transformations of a FP interferometer. Next, we go
on to predict an antibunching effect, showing a more rich
structure, due mainly to the heavy frequency dependence of the
transmissivity and reflectivity coefficients of a FP
interferometer.

This paper is organized as follows. In Section
\ref{sec:QO_Fabry_Perot} the quantum optical description of the FP
interferometer is sketched. Its equivalence to a beam splitter
having frequency-dependent transmissivity and reflectivity
coefficients is emphasized, too. The prediction of the
antibunching effect when two single-photon Fock states impinge on
the Fabry-Perot interferometer from opposite directions is
discussed in detail in Section \ref{sec:antibunching_FP_cavity}.
The extension to the case when the input light is
non-monochromatic is done in Section
\ref{sec:antibunching_FP_cavity_nonmonochrom}. The conclusions
drawn in Section \ref{sec:conclusions} close this paper.


\section{The quantum optical description of a Fabry-Perot interferometer}
\label{sec:QO_Fabry_Perot}

\begin{figure}
\centering
\includegraphics[width=3in]{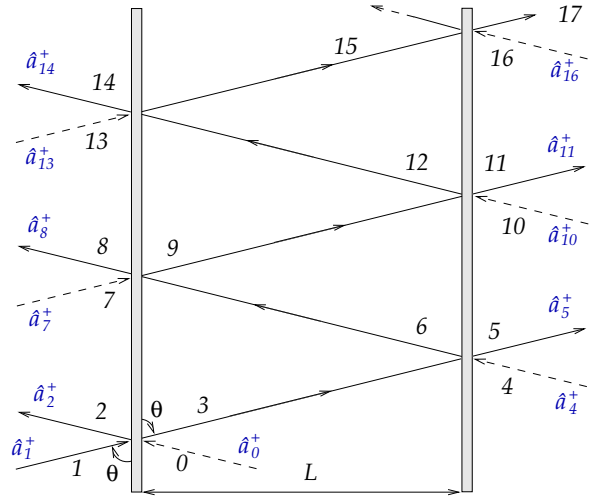}
\caption{The Fabry-Perot interferometer. Each input/output mode is
labelled with a number, starting with $0$. We associate to each
input/output mode $l$ a field creation operator denoted by
$\hat{a}^\dagger_{l}$.} \label{fig:FP_cavity_real_life}
\end{figure}

A Fabry-Perot interferometer (depicted in
Fig.~\ref{fig:FP_cavity_real_life}) is composed of two parallel
(sometimes slightly concave) highly reflecting mirrors placed at a
distance $L$. Throughout this paper, we assume the mirrors to be
identical\footnote{The case with non-identical mirrors was
discussed in \cite{Ley87}. Extension to this case is also possible
with the graphical method discussed in Appendix
\ref{app:sec:FP_graphical_method}.} and we restrict our analysis
to a single axis (i.e. orthogonal to the mirrors). Moreover, we
assume each highly reflecting mirror to be frequency independent
in the optical range of interest. Thus, we can model it by a beam
splitter (BS) having a transmissivity $T$ and a reflectivity $R$.
Since we assume our mirrors to be of negligible thickness compared
to the distance $L$, we take the transmissivity to be a positive
real number\footnote{For high reflectivity FP cavity mirrors,
$\varepsilon$ is a small real number. However, in all computations
that follow, this condition is not needed. 
} i.e.
$T=\varepsilon$ and the reflectivity to be purely imaginary,
therefore, we have
\begin{equation}
\label{eq:T_R_versus_epsilon} T=\varepsilon\quad\text{ and }\quad
R=i\sqrt{1-\varepsilon^2}
\end{equation}
One can easily check that these coefficients obey the well-known
relations for a beam splitter $\vert{T}\vert^2+\vert{R}\vert^2=1$
and $TR^*+T^*R=0$ \cite{Lou03}. The angle of incidence of our
light beam of interest is denoted by $\theta$ (see
Fig.~\ref{fig:FP_cavity_real_life}).

The FP interferometer is resonant at wavelengths $\lambda_0$ where
the round trip in the cavity of length $L$ causes a phase shift
$\varphi=kL$ that is an integer multiple of $2\pi$. Therefore, for
$\theta\to\pi/2$  we have the relation
$L=N\lambda_0/2=\pi{N}c/\omega_0$, where $\omega_0$ is the
resonant (angular) frequency of the cavity, $N\in\mathbb{N}$ and
$c$ is the speed of light in vacuum. Since $k=\omega/c$ one gets
the phase shift $\varphi=\pi{N}\omega/\omega_0=\omega t_{0}$ where
we denoted $t_{0}=L/c$. This expression will be used later on.

Since the interesting case from a practical point of view is when
the angle of incidence is almost normal ($\theta\to\pi/2$), we
shall consider this scenario for the rest of the paper. In this
case, our FP interferometer can be considered as having only four
modes (ports) as depicted in
Fig.~\ref{fig:FP_cavity_real_life_operators} (see also the
discussion in Appendix \ref{app:sec:FP_graphical_method}). We can
find the input-output operator transformations by various means.
In Appendix \ref{app:sec:FP_graphical_method}, they are obtained
via a graphical method. The end results are
\begin{equation}
\label{eq:FP_operator_transf_a_I_fct_a_T_a_R}
\hat{a}_I^\dagger=T_{fp}\hat{a}_T^\dagger+R_{fp}\hat{a}_R^\dagger
\end{equation}
and
\begin{equation}
\label{eq:FP_operator_transf_a_V_fct_a_T_a_R}
\hat{a}_V^\dagger=R_{fp}\hat{a}_T^\dagger+T_{fp}\hat{a}_R^\dagger
\end{equation}
where $T_{fp}$ and, respectively, $R_{fp}$ are given by equations
\eqref{eq:FP_T_fp_phi} and, respectively, \eqref{eq:FP_R_fp_phi}.

\begin{figure}
\centering
\includegraphics[width=2in]{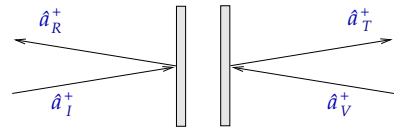}
\caption{The creation field operators for a Fabry-Perot
interferometer when $\theta\to\pi/2$. Input modes are labelled by
$I$ and $V$, while output modes are labelled by $T$ and $R$. Each
input/output mode has the corresponding creation operator attached
to it.} \label{fig:FP_cavity_real_life_operators}
\end{figure}


By analyzing equations
\eqref{eq:FP_operator_transf_a_I_fct_a_T_a_R} and
\eqref{eq:FP_operator_transf_a_V_fct_a_T_a_R} one notes that these
are actually the operator transformations of a beam splitter
having frequency dependent transmissivity
\begin{equation}
\label{eq:FP_T_transmissivity}
T_{fp}\left(\omega\right)=\frac{T^2\text{e}^{i\omega
t_0}}{1-R^2\text{e}^{i2\omega t_0}}
\end{equation}
and reflectivity
\begin{equation}
\label{eq:FP_R_reflectivity} R_{fp}\left(\omega\right)
=R\frac{1+\text{e}^{i2\omega t_0}}{1-R^2\text{e}^{i2\omega t_0}}
\end{equation}
coefficients. The probability of transmission and, respectively,
reflection of a monochromatic beam of light of frequency $\omega$
is easily computed as
\begin{equation}
\label{eq:FP_T_mod_squared}
\vert{T_{fp}\left(\omega\right)}\vert^2=\frac{1}
{1+\mathcal{F}^2\cos^2\left(\omega t_0\right)}
\end{equation}
and
\begin{equation}
\label{eq:FP_R_mod_squared}
\vert{R_{fp}\left(\omega\right)}\vert^2=\frac{\mathcal{F}^2\cos^2\left(\omega
t_0\right)} {1+\mathcal{F}^2\cos^2\left(\omega t_0\right)}
\end{equation}
where we replaced $T$ and $R$ according to equation
\eqref{eq:T_R_versus_epsilon} and the cavity finesse was defined
as $\mathcal{F}=2\sqrt{1-\varepsilon^2}/\varepsilon^2$. We note
that we have
$\vert{T_{fp}\left(\omega\right)}\vert^2+\vert{R_{fp}\left(\omega\right)}\vert^2=1$
and
$T_{fp}\left(\omega\right)R_{fp}^*\left(\omega\right)+T_{fp}^*\left(\omega\right)R_{fp}\left(\omega\right)=0$
i.e. we have a $SU(2)$ symmetry, characteristic of linear lossless
optical systems \cite{Yur86,Cam89}.

\begin{figure}
\centering
\includegraphics[width=1.2in]{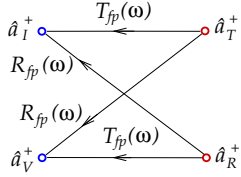}
\caption{The Fabry-Perot interferometer can be also seen as a beam
splitter with frequency-dependent $T_{fp}\left(\omega\right)$ and
$R_{fp}\left(\omega\right)$ (transmission and reflection
coefficients).} \label{fig:graph_FP_cavity_BS_equiv}
\end{figure}

Therefore, the whole Fabry-Perot interferometer can be modelled as
a simple beam splitter (see
Fig.~\ref{fig:graph_FP_cavity_BS_equiv}), with its transmissivity
and reflectivity coefficients being heavily frequency dependent.

It is then expected that, similar to a beam splitter, the same
classical and non-classical effects from quantum optics must be
reproduced by a FP interferometer.

\section{Antibunching with a Fabry-Perot interferometer in the case with monochromatic photons}
\label{sec:antibunching_FP_cavity} The antibunching is a
non-classical effect observed when two single-photon Fock states
impinge simultaneously on the two inputs of a balanced ($50/50$)
beam splitter \cite{HOM87}.

Since the Fabry-Perot interferometer is formally equivalent to a
beam splitter (with frequency dependent transmissivity and
reflectivity, as discussed before), using the $\vert1_I1_V\rangle$
input state should show this nonclassical effect at its output.
Therefore, we apply the input state to a FP interferometer
\begin{eqnarray}
\vert\psi_{in}\rangle=\vert1_I1_V\rangle=\hat{a}_I^\dagger\hat{a}_V^\dagger\vert0\rangle
\end{eqnarray}
where both input light quanta are supposed to be monochromatic of
frequency $\omega$. Using again the input-output field operator
transformation relations
\eqref{eq:FP_operator_transf_a_I_fct_a_T_a_R} and
\eqref{eq:FP_operator_transf_a_V_fct_a_T_a_R} takes us to the
output state vector
\begin{eqnarray}
\label{eq:FP_output_11_input}
\vert\psi_{out}\rangle=\left(T_{fp}\hat{a}_T^\dagger+R_{fp}\hat{a}_R^\dagger\right)
\left(R_{fp}\hat{a}_T^\dagger+T_{fp}\hat{a}_R^\dagger\right)\vert0\rangle
\nonumber\\
=T_{fp}R_{fp}\left(\vert2_T0_R\rangle+\vert0_T2_R\rangle\right)+\left(T_{fp}^2+R_{fp}^2\right)\vert1_T1_R\rangle
\end{eqnarray}
We shall denote by
$c_{HOM}\left(\varepsilon,\omega\right)=T_{fp}^2\left(\omega\right)+R_{fp}^2\left(\omega\right)$
and
$P_{HOM}\left(\varepsilon,\omega\right)=\vert{c_{HOM}\left(\varepsilon,\omega\right)}\vert^2$
the amplitude, and, respectively, the probability of coincidence
detection at the outputs $T$ and $R$ of the FP interferometer.

In order to obtain the antibunching effect, we impose the
amplitude of the $\vert1_T1_R\rangle$ state to be zero. The
condition $c_{HOM}\left(\omega\right)=0$ implies
$\cos^2\left(\omega
t_0\right)=\varepsilon^4/4\left(1-\varepsilon^2\right)$, an
equation that can be readily solved yielding
\begin{eqnarray}
\label{eq:solution_omega_vs_eps_HOM}
\omega_{+/-}=\frac{1}{t_0}\arccos\left(\pm\frac{\varepsilon^2}{2\sqrt{1-\varepsilon^2}}\right)
\end{eqnarray}
This equation has solutions as long as
$\varepsilon\leq\varepsilon_0$, where
$\varepsilon_0=\sqrt{2\left(\sqrt{2}-1\right)}$. We could see
$\varepsilon_0$ as the maximum value of the transmissivity of the
individual mirrors, so that the FP interferometer can still
exhibit the antibunching effect\footnote{It is interesting to note
that at the frequency $\omega_+$ given by equation
\eqref{eq:solution_omega_vs_eps_HOM}, we have
$T_{fp}\left(\omega_+\right)=1/2\left(1+i\right)$ and
$R_{fp}\left(\omega_+\right)=1/2\left(-1+i\right)$ while at the
frequency $\omega_{-}$ we have
$T_{fp}\left(\omega_-\right)=-1/2+i/2$ and
$R_{fp}\left(\omega_-\right)=1/2+i/2$.}.

\begin{figure}
\centering
\includegraphics[width=3.5in]{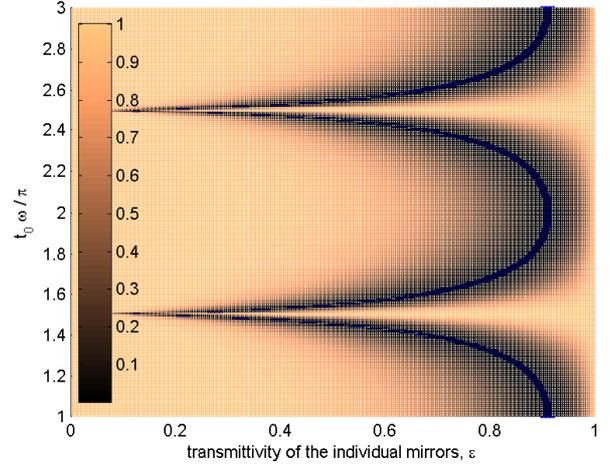}
\caption{The antibunching factor $P_{HOM}=\vert{c_{HOM}}\vert^2$
versus the transmissivity of individual mirrors $\varepsilon$ and
the phase variation $\omega t_0/\pi$. The blue thick curve
corresponds to pairs $(\varepsilon,\omega)$ yielding
$P_{HOM}(\varepsilon,\omega)\leq0.01$.}
\label{fig:Matlab_FP_antibunching_P_HOM_versus_T_omega}
\end{figure}

In practice, however, having
$P_{HOM}\left(\varepsilon,\omega\right)$ below a certain value
would be still satisfactory. In
Fig.~\ref{fig:Matlab_FP_antibunching_P_HOM_versus_T_omega} we plot
$P_{HOM}(\varepsilon,\omega)$ in respect with the transmissivity
of individual mirrors ($\varepsilon$) and the phase variation
($\omega t_0/\pi$). The thick blue curve from
Fig.~\ref{fig:Matlab_FP_antibunching_P_HOM_versus_T_omega}
outlines the values $\varepsilon$ and $\omega$ for $P_{HOM}<0.01$.

Up until now, these results are equivalent to the ones obtained
with a balanced beam splitter. The true interest in using a FP
interferometer lies in the fact that the same effect can be
obtained using photons of different colors.

\begin{figure}
\centering
\includegraphics[width=3.5in]{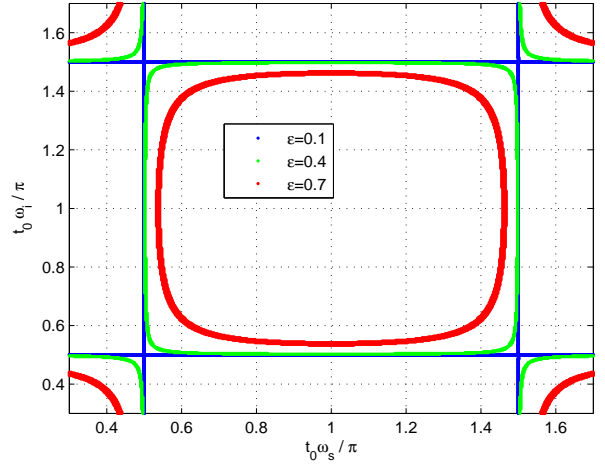}
\caption{The pairs of frequencies $\left(\omega_s,\omega_i\right)$
satisfying $P_{HOM}\left(\omega_s,\omega_i\right)<0.01$. The thick
red curve corresponds to $\varepsilon=0.7$, the green one
corresponds to $\varepsilon=0.4$ and the blue curve corresponds to
$\varepsilon=0.1$.}
\label{fig:Matlab_FP_antibunching_surface_P_HOM_versus_omega_s_i}
\end{figure}

We consider now two (monochromatic) photons of two different
frequencies, $\omega_s$ and $\omega_i$. Therefore, the input state
can be written as
\begin{eqnarray}
\label{eq:FP_psi_input_11_2color}
\vert\psi_{in}\rangle=\iint{\text{d}\omega\text{d}\omega'\zeta_I\left(\omega\right)\zeta_{\:V}\left(\omega'\right)
\hat{a}_I^\dagger\left(\omega\right)\hat{a}_V^\dagger\left(\omega'\right)}\vert0\rangle
\end{eqnarray}
with
$\zeta_I\left(\omega\right)=\delta\left(\omega-\omega_s\right)$
and
$\zeta_{\:V}\left(\omega\right)=\delta\left(\omega-\omega_i\right)$.
The frequency-dependent input-output creation operators now yield
\begin{eqnarray}
\label{eq:FP_a_I_versus_a_T_a_R_at_omega_s}
\hat{a}_I^\dagger\left(\omega_s\right)=T_{fp}\left(\omega_s\right)\hat{a}_T^\dagger\left(\omega_s\right)
+R_{fp}\left(\omega_s\right)\hat{a}_R^\dagger\left(\omega_s\right)
\end{eqnarray}
and
\begin{eqnarray}
\label{eq:FP_a_R_versus_a_T_a_R_at_omega_i}
\hat{a}_V^\dagger\left(\omega_i\right)=R_{fp}\left(\omega_i\right)\hat{a}_T^\dagger\left(\omega_i\right)
+T_{fp}\left(\omega_i\right)\hat{a}_R^\dagger\left(\omega_i\right)
\end{eqnarray}
The output state vector is obtained by considering the field
operator transformations
(\ref{eq:FP_a_I_versus_a_T_a_R_at_omega_s}$-$\ref{eq:FP_a_R_versus_a_T_a_R_at_omega_i})
on $\vert\psi_{in}\rangle$ from equation
\eqref{eq:FP_psi_input_11_2color} and gives
\begin{eqnarray}
\label{eq:FP_psi_output_11_2color} \vert\psi_{out}\rangle=
T_{fp}\left(\omega_s\right)R_{fp}\left(\omega_i\right)
\hat{a}_T^\dagger\left(\omega_s\right)\hat{a}_T^\dagger\left(\omega_i\right)\vert0\rangle+
\nonumber\\
T_{fp}\left(\omega_i\right)R_{fp}\left(\omega_s\right)
\hat{a}_R^\dagger\left(\omega_s\right)\hat{a}_R^\dagger\left(\omega_i\right)\vert0\rangle+
\nonumber\\
\left(T_{fp}\left(\omega_s\right)T_{fp}\left(\omega_i\right)
\hat{a}_T^\dagger\left(\omega_s\right)\hat{a}_R^\dagger\left(\omega_i\right)\right.
\nonumber\\
\left.+R_{fp}\left(\omega_s\right)R_{fp}\left(\omega_i\right)
\hat{a}_T^\dagger\left(\omega_i\right)\hat{a}_R^\dagger\left(\omega_s\right)\right)\vert0\rangle
\end{eqnarray}
Since the photon counters are typically frequency non-selective
(at least in a small spectral range), a coincidence detection at
the $T$ and $R$ outputs can be modelled as
\begin{eqnarray}
c_{HOM}\left(\omega_s,\omega_i\right)=\langle0\vert\iint{\text{d}\omega\text{d}\omega'
\hat{a}_T\left(\omega\right)\hat{a}_R\left(\omega'\right)}\vert\psi_{out}\rangle
\nonumber\\
=T_{fp}\left(\omega_s\right)T_{fp}\left(\omega_i\right)+R_{fp}\left(\omega_s\right)R_{fp}\left(\omega_i\right)
\end{eqnarray}
where we used the commutation relations \cite{Lou03}
\begin{eqnarray}
\label{eq:a_a_dagger_commutator} [\hat{a}_l\left(\omega\right),
\hat{a}_l^\dagger\left(\omega'\right)]=\delta\left(\omega-\omega'\right)
\end{eqnarray}
for $l\in\{T,R\}$. The antibunching condition now reads
$c_{HOM}\left(\omega_s,\omega_i\right)=0$, leading to the
constraint on $\omega_s$ and $\omega_i$:
\begin{eqnarray}
\label{eq:antibunch_eq_monochrom_omega_s_i} \cos\left(\omega_s
t_0\right)\cos\left(\omega_i
t_0\right)=\frac{\varepsilon^4}{4\left(1-\varepsilon^2\right)}
\end{eqnarray}
Again, in a practical scenario, we might relax our constraints
and, instead of imposing a zero amplitude for
$c_{HOM}\left(\omega_s,\omega_i\right)$, we could ask for a value
of $P_{HOM}\left(\omega_s,\omega_i\right)=\vert
c_{HOM}\left(\omega_s,\omega_i\right)\vert^2$ below a certain
limit. The points $\left(\omega_s,\omega_i\right)$ satisfying
$P_{HOM}\left(\omega_s,\omega_i\right)<0.01$ are plotted in
Fig.~\ref{fig:Matlab_FP_antibunching_surface_P_HOM_versus_omega_s_i}
for three different values of $\varepsilon$.

In Appendix \ref{app:sec:antibunching_monochrom_SPDC} we discuss
what changes implies the more realistic assumption that the signal
and idler photons originate from a SPDC process.

\section{Antibunching with a Fabry-Perot interferometer in the case of non-monochromatic photons}
\label{sec:antibunching_FP_cavity_nonmonochrom} We extend now the
previous analysis to non-monochromatic photon wave packets
\cite{Blo90,Cam90}. The input state can be expressed as
\begin{eqnarray}
\vert\psi_{in}\rangle=\iint{\text{d}\omega\text{d}\omega'
\zeta\left(\omega,\omega'\right)\hat{a}_I^\dagger\left(\omega\right)\hat{a}_V^\dagger\left(\omega'\right)}\vert0\rangle
\end{eqnarray}
where $\zeta\left(\omega,\omega'\right)$ is the bi-photon wave
packet and it is normalized via
$\iint{\vert\zeta\left(\omega,\omega'\right)\vert^2\text{d}\omega\text{d}\omega'}=1$.
The output positive frequency electric field operators have to be
extended to
\begin{eqnarray}
\label{eq:E_positive_freq_non_monochrom_output}
\hat{E}_{l}^{(+)}\left(t\right)=\frac{1}{\sqrt{2\pi}}\int{\text{d}\omega\text{e}^{-i\omega{t}}\hat{a}_l\left(\omega\right)}
\end{eqnarray}
where $l\in\{T,R\}$. We shall be interested in the second order
correlation function \cite{Lou03,Man95}
\begin{eqnarray}
\label{eq:G_2_t_tau_first}
G^{(2)}\left(t,t+\tau\right)=
\langle\psi_{out}\vert\hat{E}_{T}^{(-)}\left(t\right)\hat{E}_{R}^{(-)}\left(t+\tau\right)
\nonumber\\
\hat{E}_{R}^{(+)}\left(t+\tau\right)\hat{E}_{T}^{(+)}\left(t\right)
\vert\psi_{out}\rangle
\end{eqnarray}
where
$\hat{E}_{l}^{(-)}\left(t\right)=[\hat{E}_{l}^{(+)}\left(t\right)]^\dagger$
and $t$ ($t+\tau$) denotes the detection time at the output $T$
($R$). We will use the Schr\"odinger picture therefore we shall
evolve $\vert\psi_{in}\rangle$ to $\vert\psi_{out}\rangle$. Using
again the field operator transformations
\eqref{eq:FP_operator_transf_a_I_fct_a_T_a_R} and
\eqref{eq:FP_operator_transf_a_V_fct_a_T_a_R} obtaining the output
state vector
\begin{eqnarray}
\label{eq:Psi_out_non_monochrom_11_state}
\vert\psi_{out}\rangle=\iint{\text{d}\omega\text{d}\omega'
}\left(T_{fp}\left(\omega\right)\hat{a}_T^\dagger\left(\omega\right)
+R_{fp}\left(\omega\right)
\hat{a}_R^\dagger\left(\omega\right)\right)
\nonumber\\
\times\left(R_{fp}\left(\omega'\right)
\hat{a}_T^\dagger\left(\omega'\right)+T_{fp}\left(\omega'\right)
\hat{a}_R^\dagger\left(\omega'\right)\right)\zeta\left(\omega,\omega'\right)\vert0\rangle\quad
\end{eqnarray}
Combining now equations
\eqref{eq:E_positive_freq_non_monochrom_output},
\eqref{eq:G_2_t_tau_first} and
\eqref{eq:Psi_out_non_monochrom_11_state}, after some
computations, one ends up with\footnote{Typically, for the
antibunching effect we shall be interested in
$G^{(2)}\left(\tau\right)$ which is the time integrated version of
$G^{(2)}\left(t,t+\tau\right)$ i.e.
$G^{(2)}\left(\tau\right)=\int{G^{(2)}\left(t,t+\tau\right)\text{d}t}$.}
\begin{eqnarray}
\label{G_2_of_0_non_monochtom_general}
G^{(2)}\left(t,t+\tau\right)=\frac{1}{2\pi}\bigg\vert
\iint{\text{d}\omega\text{d}\omega'\text{e}^{-i\omega{t}}\text{e}^{-i\omega'\left(t+\tau\right)}
}\qquad
 \nonumber\\
\times\left(T_{fp}\left(\omega\right)T_{fp}\left(\omega'\right)
+R_{fp}\left(\omega\right)R_{fp}\left(\omega'\right)\right)
\zeta\left(\omega,\omega'\right)\bigg\vert^2
\end{eqnarray}
where we used again the commutation relations
\eqref{eq:a_a_dagger_commutator}. If we make the simplifying
assumption that the signal and idler frequencies are not frequency
correlated i.e. we have
$\zeta\left(\omega,\omega'\right)=\zeta_0\left(\omega\right)\zeta_1\left(\omega'\right)$,
equation \eqref{G_2_of_0_non_monochtom_general} simplifies to
\begin{eqnarray}
G^{(2)}\left(t,t+\tau\right)=\vert
\zeta_{0}^{T}\left(t\right)\zeta_{1}^{T}\left(t+\tau\right)
+\zeta_{0}^{R}\left(t+\tau\right)\zeta_{1}^{R}\left(t\right)\vert^2\quad
\end{eqnarray}
where we have the inverse Fourier transforms
\begin{eqnarray}
\zeta_{m}^{T}\left(t\right)=\frac{1}{\sqrt{2\pi}}\int{\text{d}\omega\text{e}^{-i\omega{t}}T_{fp}\left(\omega\right)\zeta_m\left(\omega\right)}
\end{eqnarray}
and
\begin{eqnarray}
\zeta_{m}^{R}\left(t\right)=\frac{1}{\sqrt{2\pi}}\int{\text{d}\omega\text{e}^{-i\omega{t}}R_{fp}\left(\omega\right)\zeta_m\left(\omega\right)}
\end{eqnarray}
with $m\in\{0,1\}$. To the author's best knowledge, there is no
closed-form solution to these Fourier integrals. However, one can
easily check that if
$\zeta_0\left(\omega\right)\to\delta\left(\omega-\omega_s\right)$
and
$\zeta_1\left(\omega'\right)\to\delta\left(\omega'-\omega_i\right)$
equation \eqref{G_2_of_0_non_monochtom_general} migrates into
\begin{eqnarray}
G^{(2)}\sim\vert
T_{fp}\left(\omega_s\right)T_{fp}\left(\omega_i\right)
+R_{fp}\left(\omega_s\right)R_{fp}\left(\omega_i\right)\vert^2
\end{eqnarray}
If $\omega_s$ and $\omega_i$ obey equation
\eqref{eq:antibunch_eq_monochrom_omega_s_i}, then the antibunching
effect is assured. If one assumes the SPDC-like frequency
correlation,
$\zeta\left(\omega,\omega'\right)\to\delta\left(\omega-\omega_s\right)\delta\left(\omega'-\omega_p+\omega_s\right)$
the discussion from Appendix
\ref{app:sec:antibunching_monochrom_SPDC} still holds.

\section{Conclusions}
\label{sec:conclusions}

In this paper we started from the full quantum optical description
of a Fabry-Perot interferometer and showed that it can be modelled
through a beam splitter having heavily frequency-dependent
transmission and reflection coefficients.

Owing to this model, the antibunching effect, well-known for beam
splitters has been predicted for the Fabry-Perot interferometer,
too. However, even different frequency photons can show this
effect, this being in contrast with the normal antibunching.
Monochromatic, as well as non-monochromatic input light was
considered.


\appendix

\section{The computation of the field operator transformations in a FP interferometer}
\label{app:sec:FP_graphical_method}

\begin{figure*}
\centering
\includegraphics[width=6.5in]{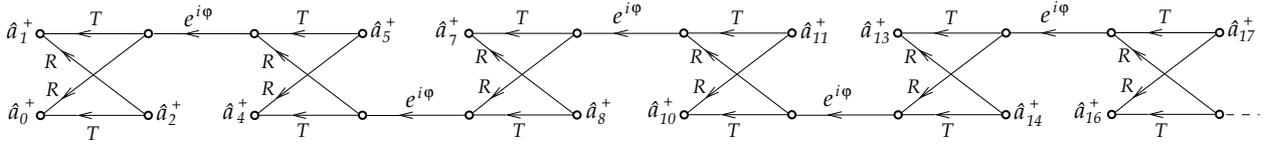}
\caption{The graphical description of a Fabry-Perot interferometer
for light incident at an arbitrary angle. Each passage of the beam
through the highly reflecting mirrors is modelled with a beam
splitter. The labelling was done in accordance to
Fig.~\ref{fig:FP_cavity_real_life}.} \label{fig:graph_FP_cavity}
\end{figure*}

In the following we will employ the graphical method introduced in
\cite{Ata14b}. Beam splitters will be depicted by the
butterfly-like structure and delays will simply add a
$\text{e}^{i\varphi}$ factor. The graphical representation of the
Fabry-Perot interferometer is detailed in
Fig.~\ref{fig:graph_FP_cavity}. Each crossing of a
highly-reflecting mirror is modelled by a beam splitter (having
the transmission and reflection coefficients $T$ and,
respectively, $R$) and each passage inside the cavity yields an
extra factor of $\text{e}^{i\varphi}$ where $\varphi=kL$ and $k$
is the wavenumber of the (monochromatic) light that illuminates
the interferometer.

It is not difficult to find the expression of the input creation
operator $\hat{a}_1^\dagger$ in respect with the output ones. We
can differentiate between transmission type creation operators
($\hat{a}_{5}^\dagger$, $\hat{a}_{11}^\dagger$,
$\hat{a}_{17}^\dagger$ $\ldots$) and reflection type operators
($\hat{a}_2^\dagger$, $\hat{a}_8^\dagger$, $\hat{a}_{14}^\dagger$
$\ldots$). Adding up all these contributions, one ends up with
\begin{eqnarray}
\label{eq:FP_operator_transf_series}
\hat{a}_1^\dagger=T^2\text{e}^{i\varphi}
\left(\hat{a}_{5}^\dagger+R^2\text{e}^{i2\varphi}\hat{a}_{11}^\dagger
+R^4\text{e}^{i4\varphi}\hat{a}_{17}^\dagger+\ldots\right)
\nonumber\\
+R\left(\hat{a}_2^\dagger+T^2\text{e}^{i2\varphi}\hat{a}_8^\dagger
+T^2R^2\text{e}^{i4\varphi}\hat{a}_{14}^\dagger+\ldots\right)
\end{eqnarray}
As discussed before, the interesting case from a practical point
of view is when the angle $\theta$ tends to $\pi/2$. The many
inputs and outputs tend to group together in the following way:
the inputs ($\hat{a}_1^\dagger$, $\hat{a}_7^\dagger$,
$\hat{a}_{13}^\dagger$ $\ldots$) become
indistinguishable\footnote{Although one might wish to input a
photon only at the port $1$, the extreme closeness of the ports
$1$, $7$, $13$ $\ldots$ makes impossible to tell which port
actually received the photon.} and we relabel them as a single
input $I$, whose creation operator will be denoted by
$\hat{a}_I^\dagger$. Similarly the reflected outputs
($\hat{a}_2^\dagger$, $\hat{a}_8^\dagger$, $\hat{a}_{14}^\dagger$
$\ldots$) become also indistinguishable and will be denoted by
$\hat{a}_R^\dagger$. The same logic applies in regrouping the
appropriate transmission operators ($\hat{a}_5^\dagger$,
$\hat{a}_8$, $\hat{a}_{11}^\dagger$, $\hat{a}_{17}^\dagger$
$\ldots$) into $\hat{a}_T^\dagger$. The ``dark'' ports
($\hat{a}_0^\dagger$, $\hat{a}_4^\dagger$, $\hat{a}_{10}^\dagger$,
$\hat{a}_{16}^\dagger$ $\ldots$) are grouped into the ``vacuum''
port $\hat{a}_V^\dagger$ (see
Fig.~\ref{fig:FP_cavity_real_life_operators}). Now equation
\eqref{eq:FP_operator_transf_series} takes the much simpler form
\begin{eqnarray}
\label{eq:FP_operator_transf_series2}
\hat{a}_I^\dagger=T^2\text{e}^{i\varphi}
\left(1+R^2\text{e}^{i2\varphi}+R^4\text{e}^{i4\varphi}+\ldots\right)\hat{a}_{T}^\dagger
\nonumber\\
+R\left(1+T^2\text{e}^{i2\varphi}+T^2R^2\text{e}^{i4\varphi}+\ldots\right)\hat{a}_{R}^\dagger
\end{eqnarray}
Since we have now two trivial geometric series and
$\vert{R}\vert<1$, equation \eqref{eq:FP_operator_transf_series2}
can be written in the rather simple form
\begin{equation}
\hat{a}_I^\dagger=\frac{T^2\text{e}^{i\varphi}}{1-R^2\text{e}^{i2\varphi}}\hat{a}_T^\dagger+
R\left(1+\frac{T^2\text{e}^{i2\varphi}}{1-R^2\text{e}^{i2\varphi}}\right)\hat{a}_R^\dagger
\end{equation}
We could define the transmission
\begin{equation}
\label{eq:FP_T_fp_phi}
T_{fp}=\frac{T^2\text{e}^{i\varphi}}{1-R^2\text{e}^{i2\varphi}}
\end{equation}
and reflection
\begin{equation}
\label{eq:FP_R_fp_phi} R_{fp}=
R\left(1+\frac{T^2\text{e}^{i2\varphi}}{1-R^2\text{e}^{i2\varphi}}\right)
\end{equation}
coefficients of the FP interferometer allowing the easy writing of
the field operator transformation given by equation
\eqref{eq:FP_operator_transf_a_I_fct_a_T_a_R}.
%
Similar arguments allow one to write $\hat{a}_V^\dagger$ in
respect with $\hat{a}_T^\dagger$ and $\hat{a}_R^\dagger$, as done
in equation \eqref{eq:FP_operator_transf_a_V_fct_a_T_a_R}.

\begin{figure}
\centering
\includegraphics[width=2.5in]{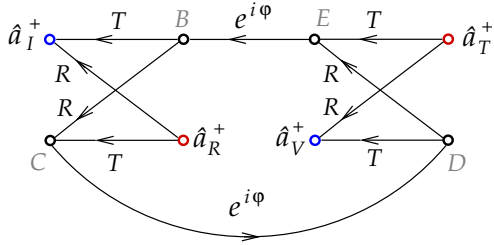}
\caption{The graphical description of a Fabry-Perot interferometer
in the recursive form when light impinges at a normal angle. The
blue circles denote the input ports while the red ones denote the
output ports. Every path having a common segment with the loop
$B-C-D-E$ gains a factor of $1/(1-R^2\text{e}^{i2\varphi})$.}
\label{fig:graph_FP_cavity_recursive}
\end{figure}

The regrouping of input and output field operators that became
indistinguishable allows a much simpler graphical representation
of the FP interferometer, as depicted in
Fig.~\ref{fig:graph_FP_cavity_recursive}.

The infinite series of beam splitters from
Fig.~\ref{fig:graph_FP_cavity} was replaced by a loop, having the
nodes $B$, $C$, $D$ and $E$ in
Fig.~\ref{fig:graph_FP_cavity_recursive}. The meaning of the loop
is that besides the direct path, there is a possibility to loop
ones, twice and so forth. For example, in
Fig.~\ref{fig:graph_FP_cavity_recursive}, the direct path from
$\hat{a}_T^\dagger$ to $\hat{a}_I^\dagger$ took simply a factor of
$\text{e}^{i\varphi}$ (besides the $T^2$ factor caused by elements
outside of the loop). Looping ones implies, besides
$\text{e}^{i\varphi}$, a supplementary factor of
$R^2\text{e}^{i2\varphi}$. Looping twice implies a supplementary
factor of $R^4\text{e}^{i4\varphi}$ and so forth. This is the
geometric series
($1+R^2\text{e}^{i2\varphi}+R^4\text{e}^{i4\varphi}+\ldots$) we
found before and it trivially yields
$1/(1-R^2\text{e}^{i2\varphi})$. In the end, from the recursive
graphical method we obtain
$T^2\text{e}^{i\varphi}/(1-R^2\text{e}^{i2\varphi})$, a result
identical to equation \eqref{eq:FP_T_fp_phi}.

Therefore, as a rule, in the graph from
Fig.~\ref{fig:graph_FP_cavity_recursive}, every input-output path
having a common segment with this loop must take a supplementary
factor of $1/(1-R^2\text{e}^{i2\varphi})$.

For example, the input creation operator $\hat{a}_I^\dagger$ can
be reached from the output port $\hat{a}_R^\dagger$ through the
reflection coefficient of the input beam splitter yielding a
factor $R$ (and touching no loop). There is a second contribution
from $\hat{a}_R^\dagger$ to $\hat{a}_I^\dagger$: transmission $T$
from the first (left) beam splitter, a factor of
$\text{e}^{i\varphi}$, a $R$ (from the second i.e. right beam
splitter), another factor of $\text{e}^{i\varphi}$ and finally a
factor of $T$ to $\hat{a}_I^\dagger$. This path has common
segments with the loop, therefore it also takes the factor
$1/(1-R^2\text{e}^{i2\varphi})$. Adding up the two possible routes
gives $R+RT^2\text{e}^{i2\varphi}/(1-R^2\text{e}^{i2\varphi})$ and
we found straight away the result from equation
\eqref{eq:FP_R_fp_phi}.

\section{The antibunching condition with photons originating in a SPDC process}
\label{app:sec:antibunching_monochrom_SPDC} In the SPDC process,
the signal and idler frequencies are bound by the relation
$\omega_p=\omega_s+\omega_i$ where $\omega_p$ denotes the pump
frequency. This constraint modifies equation
\eqref{eq:antibunch_eq_monochrom_omega_s_i} and takes us to a
single variable (say $\omega_s$) yielding
\begin{eqnarray}
\alpha_p\cos\left(2\omega_s t_0\right)+
\sqrt{1-\alpha_p^2}\sin\left(2\omega_s t_0\right)
=\beta_\varepsilon-\alpha_p
\end{eqnarray}
where we denoted $\alpha_p=\cos\left(\omega_pt_0\right)$ and
$\beta_\varepsilon=\varepsilon^4/2\left(1-\varepsilon^2\right)$.
Using well-known trigonometric identities, we transform the sines
and cosines into $\tan\left(\omega_s t_0\right)$ and have
\begin{eqnarray}
\beta_\varepsilon\tan^2\left(\omega_s
t_0\right)-2\sqrt{1-\alpha_p^2}\tan\left(\omega_s t_0\right)
+\beta_\varepsilon-2\alpha_p=0\quad
\end{eqnarray}
This second order equation in $\tan\left(\omega_s t_0\right)$
yields two solutions
\begin{eqnarray}
\tan\left(\omega_s t_0\right)\vert_{+/-}=\frac{\sqrt{1-\alpha_p^2}
\pm\sqrt{1-\alpha_p^2-\beta^2_\varepsilon+2\alpha_p\beta_\varepsilon}}{\beta_\varepsilon}
\quad
\end{eqnarray}
as long as the quantity under the square root is not negative.
This constraint imposes
\begin{eqnarray}
\alpha_p^2-2\alpha_p\beta_\varepsilon+\beta^2_\varepsilon-1\leq0
\end{eqnarray}
This can happen only if $\alpha_p$ is
\begin{eqnarray}
\beta_\varepsilon-1\leq\alpha_p\leq\beta_\varepsilon+1
\end{eqnarray}
The second inequality is always satisfied since
$\beta_\varepsilon\geq0$, therefore we are left with
\begin{eqnarray}
\frac{\varepsilon^4}{2\left(1-\varepsilon^2\right)}-1\leq\cos\left(\omega_p
t_0\right)
\end{eqnarray}
This relation puts constraints on the values of $t_0$, $\omega_p$
and $\varepsilon$ in order to make possible the antibunching
effect at the signal frequency $\omega_s$.


\end{document}